\documentstyle[twocolumn,floats,aps,prb,epsfig]{revtex}

\begin{document}
\twocolumn[\hsize\textwidth\columnwidth\hsize\csname@twocolumnfalse\endcsname%
\title{ Energetics of Quantum Antidot States in Quantum Hall Regime} 
\author{I. J. Maasilta and V. J. Goldman}
\address{Department of Physics, State University of New York, Stony Brook, NY
11794-3800}
\date{Aug 26, 1997}

\maketitle
\begin{abstract} We report experiments on the energy structure of antidot-bound states. By measuring resonant tunneling line widths as function of temperature, we determine the coupling to the remote global gate voltage and find that the effects of interelectron interaction dominate. Within a simple model, we also determine the energy spacing of the antidot-bound states, self-consistent edge electric field, and edge excitation drift velocity.
\end{abstract}

\vskip2pc]

Quantum antidots in quantum Hall (QH) regime were successfully used to demonstrate charge rigidity (quantization) of a QH condensate and to measure the charge of elementary excitations, both in the integer and fractional QH effect. \cite{gold}
In these experiments  the quantized states were controlled electrostatically by a remote global gate, and therefore determination of the coupling parameter $\alpha$ between the gate voltage $V_{G}$ and the energy $E_m$ of the antidot-bound state at the chemical potential $\mu$, $\alpha= \mid d(E_m -\mu )/dV_{G}\mid $, is necessary for quantitative interpretation of the results in terms of energy. In the case of quantum dots \cite{dots} this coupling is usually discussed within a phenomenological non-interacting electron model, which gives $\alpha$ in terms of geometrical capacitances only. On the other hand, for quantum antidots no attempt to measure or model the effect of a remote gate on energetics has been reported.

In this Communication we report new experimental results on the energy structure of a quantum antidot in the quantum Hall regime. We study the energy spectrum of the antidot-bound states using the technique of thermal excitation. By measuring resonant tunneling (RT) line widths as a function of temperature, we determine the coupling constant $\alpha$ and find $\alpha_{(1)}=12 \pm 4 \ \mu$eV/V at filling factor $\nu=1$ and $\alpha_{(1/3)}=37 \pm 1 \ \mu$eV/V at $\nu=1/3$. Surprisingly, these values are equal to the value of $\alpha_0 =d\mu /dV_{G}$ obtained in the model of two-dimensional non-interacting electrons at magnetic field $B=0$: $\alpha_{(1)}$ is equal to $\alpha_0=12.2\ \mu$eV/V, and $\alpha_{(1/3)}=3\alpha_0$. That is, $\alpha_{(1/3)}$ has value of  $\alpha_0$ for charge $e/3$ particles with density of states equal to that of free spin-polarized electrons in zero $B$. This observation does not appear to be a numerical coincidence, but is not fully understood at present. Our results suggest that self-consistent electrostatics of {\em interacting} electrons forming the edge channel should play a central role in the microscopic understanding of this problem. 

\begin{figure}[hbt]
\centerline{\epsfig{file=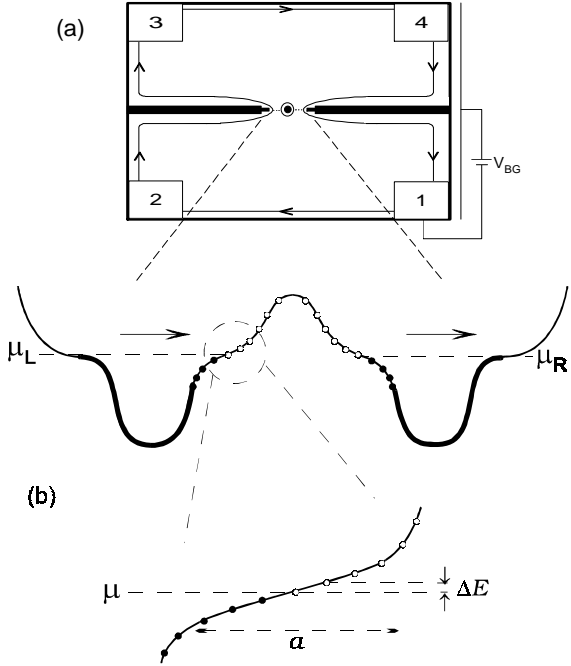,width=\linewidth,angle=0}}
\vspace{1in}
\caption{(a) Illustration of the sample. Numbered rectangles are Ohmic contacts, black areas are front gates, and arrowed lines show edge channels. The back gate extends over the entire sample on the opposite side of the insulating substrate. Dotted line represents tunneling path. (b) Schematic self-consistent energy landscape near the antidot. The ladder of quantized states around the antidot has filled ($\bullet$) and empty ($\circ$) states. Tunneling from left edge at chemical potential $\mu_L$ to the right edge at $\mu_R$ occurs through $m$-th single level if
 $\mu_L-\mu_R, k_BT \ll \Delta E = E_{m-1} - E_m$. The QH gap forms the two tunneling barriers. The blow-up shows the quantized edge channel of width $a$ circulating the antidot.} 
\label{sample}
\end{figure}

Samples were fabricated from very low disorder GaAs/AlGaAs heterosturcture material. The antidot-in-a-constriction geometry [see Fig. 1 (a)]
 was defined by standard electron beam lithography. A global back gate is separated from the 2DES by an insulating GaAs of thickness $\approx 430\ \mu$m. The two front gates were contacted independently and were used to bring the two edges close enough to the antidot for tunneling to occur.\cite{nuB} We prepared 2DES with $n\approx 1 \times 10^{11} {\rm cm^{-2}}$ and mobility $2 \times 10^{6} {\rm cm^2/Vs}$ by exposing the sample to red light at 4.2 K. Experiments were performed in a dilution refrigerator with sample probe wires filtered at mK temperatures so that the total electromagnetic background at the sample's contacts is $\sim1 \mu$V rms.  The four-terminal magnetoresistance $R_{4T}$  was measured with a lock-in amplifier. Tunneling conductance $G_T$ between the two edges can then be calculated from $R_{4T}$ as discussed previously.\cite{gold,maa}

Fig. 1(b) shows schematically the self-consistent energy landscape near the antidot in the QH regime. There is an edge channel around each of the front gates; at these edges the energy spectrum is continuous and there is no gap for charged excitations at $\mu$. Because of the electron Coulomb interaction, the self-consistent potential is flat where these edges cross $\mu$. The size of the antidot is small enough that the particle states encircling the antidot are quantized. These quantized states are the levels through which resonant tunneling occurs, and for small enough Hall voltage $\mu_L-\mu_R$ and low enough temperature $T$ tunneling takes place through a single $m$-th level at energy $E_m$. The spectrum of charged excitations is discrete in antidot-bound states even for interacting electrons, thus the self-consistent potential is not perfectly flat at $\mu$. In other words, even though the external (confining) potential is screened by 2DES in the compressible edge channels \cite{chk}, there still remains a finite slope, as shown in Fig. 1(b), when the screening length is comparable to the circumference of the channel.

\begin{figure}[t]
\centerline{\psfig{file=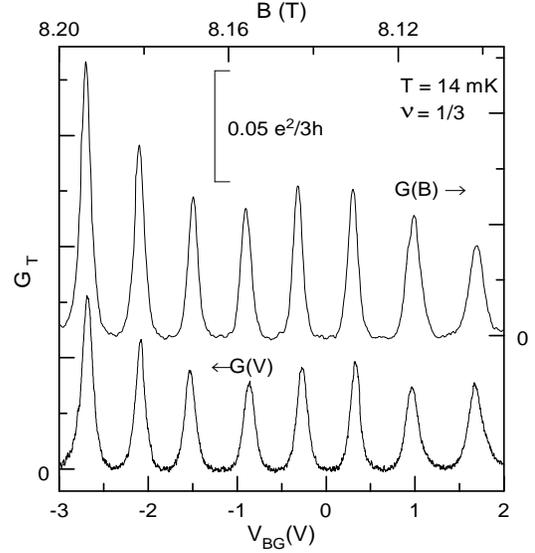,width=\linewidth,angle=0}}
\caption{Tunneling conductance of quasielectrons vs. back gate voltage and magnetic field at $\nu=1/3$.}
\label{data}
\end{figure}

The resonant level $E_m$ can be moved in energy {\em relative} to $\mu$ by changing either the magnetic field $B$ or the back gate voltage $V_{BG}$, and thus line shapes of RT peaks can be measured. Fig. 2 shows representative experimental conductance $G_T$ as function of $V_{BG}$ and $B$ at $\nu=1/3$.  We clearly observe an interval of quasiperiodic resonant tunneling peaks on top of the QH plateau, and see the equivalence between $V_{BG}$- and $B$-sweeps. As we showed in Ref. 4, thermally broadened Fermi-Dirac line shape $G_T \propto \cosh^{-2}[(E_m-\mu)/2k_BT]$ fits all the RT peaks extremely well at all temperatures.\cite{Lutt} In terms of $V_{BG}$, the line shape is written 

\begin{equation}
G_T \propto \cosh^{-2}\left [ \frac{\alpha (V_{BG}^m-V_{BG})}{2k_BT} \right ],
\label{lines}
\end{equation} 
where $\alpha= \mid d(E_m -\mu )/dV_{G}\mid $ as before, and $V_{BG}^m$ is the position of the $m$-th peak. Eq. (\ref{lines}) shows how $\alpha$ can be measured in our experiment by studying the $T$-dependence of the line shape of the RT peak.

\begin{figure}[t]
\centerline{\psfig{file=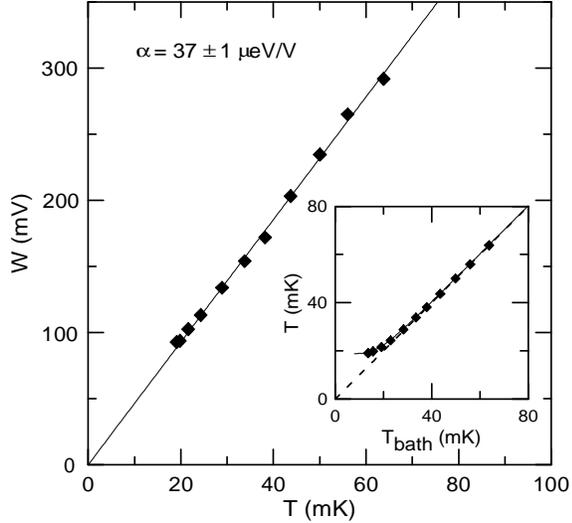,width=\linewidth,angle=0}}
\caption{Width of a quasiparticle RT peak $W$ as a function of electron temperature $T$ at $\nu=1/3$. The solid line is the one parameter linear fit giving the value for energy to voltage coupling parameter $\alpha$. The inset shows the electron $T$ as a function of the bath $T_{bath}$. The dashed line gives $T=T_{bath}$, the small deviation at lowest temperatures is accounted for by electron Joule heating, as shown by the solid line fit.}
\label{result}
\end{figure}

In Figure 3 we plot the width $W$ of a RT conductance peak, defined by $G_T \propto \cosh^{-2}[(V_{BG}^m-V_{BG})/W]$, as a function of electron temperature $T$. Comparing with Eq. (\ref{lines}), we see that $\alpha=2k_BT/W$. $W$ is directly proportional to $T$, as expected, and the slope gives the value $\alpha_{(1/3)}=37\pm 1 \ \mu$eV/V. $T$ was determined from an electron Joule heating model fit to the $W$ vs. $T_{bath}$ data, \cite{maa} where $T_{bath}$ is the LHe bath temperature, as shown in the inset in Fig. 3. Note that the deviation of $T$ from $T_{bath}$ is small and seen only at lowest temperatures. Similarly, we determine $\alpha_{(1)}=12 \pm 4 \ \mu$eV/V  for the $\nu=1$ integer QH plateau.\cite{sample}

Let us now consider  the action of a global remote back gate on a uniform two-dimensional electron system (2DES). The inverse capacitance per unit area is given by the sum of two contributions, a large geometric part and a small term related to finite compressibility of the 2DES: $1/C=1/C_{ins}+1/(e^2D_B)$. Here $C_{ins}=\epsilon \epsilon_0/d$ is the capacitance per unit area of the thickness $d$ insulator separating the 2DES from the gate, and $D_B=dn/d\mu$ is the thermodynamic density of states (DOS) at $\mu$, in a magnetic field $B$. Then  

\begin{equation}
\alpha_B=\frac{d\mu}{dV_{BG}}=\frac{d\mu}{dn}\frac{dn}{dV_{BG}}= \frac{C_B}{e}\frac{1}{D_0}+e \frac{\Delta C}{C_0},
\label{alpha}
\end{equation}
where $\Delta C=C_0-C_B$, and subscripts $0$ and $B$ refer to zero and finite magnetic fields, respectively.\cite{krav} Let us examine this result more carefully. First of all, at $B=0$ we simply get $\alpha_0=C_0/(e D_0)$. Since for non-interacting spin-polarized electrons $D_0 = m^*/(2\pi \hbar^2)$, we can compare $e^2D_0$ with $C_{ins}$. For our sample, since the back gate is {\em remote}, $e^2D_0=0.82\times 10^5  C_{ins}$. Therefore $C_0=C_{ins}[1- {\cal O}(10^{-5})]$, and to a $10^{-5}$ accuracy

\begin{equation}
\alpha_0 = \frac{C_{ins}}{e} \frac{2\pi \hbar^2}{m^*}.
\label{alpha0}
\end{equation}
Similarly, at finite values of $B$ and for particles of charge $q$, we can write 

\begin{equation}
\alpha_B = \frac{C_{ins}}{q D_B}
\label{alphaB}
\end{equation}
with error of the order of $10^{-5}$. Eq. (4) shows that $\alpha_B$ of a gated 2DES sample in a large (quantizing) $B$ is not expected to be related to the zero field result, Eq. (3), in any simple manner, since DOS will develop peaks at Landau level energies.

Comparing the experimental values of $\alpha$ to Eq. (4) we can solve for {\em quasiparticle} $D_{(1/3)}$ by using $q=e/3$. Surprisingly, we get $D_{(1/3)} \approx D_0$ ($\alpha_0=36.7 \ \mu$eV/V for $e/3$ particles), that is, it appears as if the DOS for quasiparticle excitations in the QH edge equals to the DOS of spin-polarized non-interacting electrons at $B=0$. We note that this result is distinct from the charge quantization reported in Ref. 1: the charge quantization is implied from the periodicity of the conductance peak positions in $V_{BG}$, while here we use one $\delta$-function like antidot-bound state to measure the thermal excitation spectrum on the edges {\em within} one conductance peak. \cite{bosu}

We have also analyzed RT conductance peaks at $\nu=1$ and at $\nu=1/3$  at different magnetic fields (different {\em bulk} filling factors \cite{nuB}). This data was taken at constant $T$, which is justified by the fact that phenomenologically all peaks are well described by Fermi-Dirac distribution \cite{maa}. However, so determined $\alpha$ is less accurate. In addition, the coupling parameter $\beta= \mid d(E_m -\mu )/dB\mid $ was determined from the $B$-sweep data: $G_{T} \propto \cosh^{-2}(\beta B/2k_BT)$. These results are summarized in Fig. 4. As can be seen in Fig. 4(a), $\alpha_{(1/3)}$ is approximately constant in the range $8 < B < 12$ T (within the experimental uncertainty), and agrees with the one obtained in the more accurate $T$-dependent analysis, Fig. 3. We obtain the value of $\alpha_{(1)}$ approximately three times less than $\alpha_{(1/3)}$, independent of magnetic field, while the coupling $\beta$ stays approximately constant, Fig. 4(b). In fact, the value of the ratio $\alpha/\beta$ approximately equals $\phi_0 C_{ins}/q$, as shown in Fig. 4(c). Again, the values of $\alpha$ and $\beta$ were obtained from the line shape data of a single RT peak, in contrast to the measurement of the charge of tunneling particles, which was obtained from the periods $\Delta B$ and $\Delta V_{BG}$ between consecutive RT peaks.\cite{gold,Ref 10} 

\begin{figure}[t]
\centerline{\psfig{file=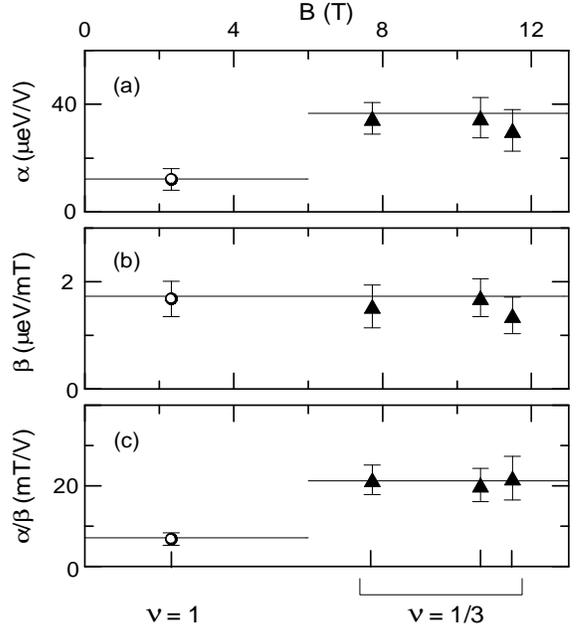,width=\linewidth,angle=0}}
\caption{(a) The energy to back-gate-voltage coupling $\alpha$, (b) energy to magnetic-field coupling $\beta$, and (c) their ratio as a function of 
magnetic field. The horizontal lines in (a) give the values of $\alpha_0$ for $\nu=1$, and 3$\alpha_0$ for $\nu={1/3}$. The line in (b) gives $\beta= 1/\phi_0 D_0$.  The horizontal lines in (c) give the expected value of $\alpha /\beta =C_{ins}\phi_0/q$. }
\label{albe}
\end{figure}

This nontrivial result, that $\alpha_{(1/3)} \approx 3\alpha_{(1)} \approx 3\alpha_0$, is not fully understood at present. The only plausible explanation in the {\em absence} of electron Coulomb interactions would seem to be loss of Landau quantization, which is ruled out by the simple fact that we do observe QH effect. On the other hand, for interacting electrons $D_B$ at $\mu$ depends sensitively on the self-consistent edge electrostatics. In an edge channel (Fig. 1), the self-consistent radial electric field ${\cal E}$ is small but finite. $\cal E$ can change as a function of $B$ to produce a particular value for $D_B$, and therefore for $\alpha$. In fact, a handwaving argument can be made that action of a {\em remote} gate on a 2DES, when $eV_{BG} >> \mu$ and all other relevant energies in the problem, should be nearly independent of an applied magnetic field and presence or absence of the QH effect. 

The values of $\alpha$ and $\beta$ can be used to obtain several properties of the antidot-bound states within the linearized ${\cal E}$ model depicted in Fig. 1 (b). The energy separation between the two consecutive states at $\mu$,  $\Delta E = E_{m-1} - E_m$ is obtained from either $\beta \Delta B$ or $\alpha \Delta V_{BG}$. Fig. 5(a) shows these results for $\Delta E$ at $\nu=1$ and at $\nu=1/3$ with different bulk filling factors.\cite{nuB} $\Delta E \approx 20 \ \mu$eV remains approximately constant as a function of $B$, even when we go from integer to fractional QH regime. This is different from the $\Delta E \propto 1/B$ dependence expected for an antidot with strong confinement potential, and as was seen in other experiments.\cite{sach} 

\begin{figure}[t]
\centerline{\psfig{file=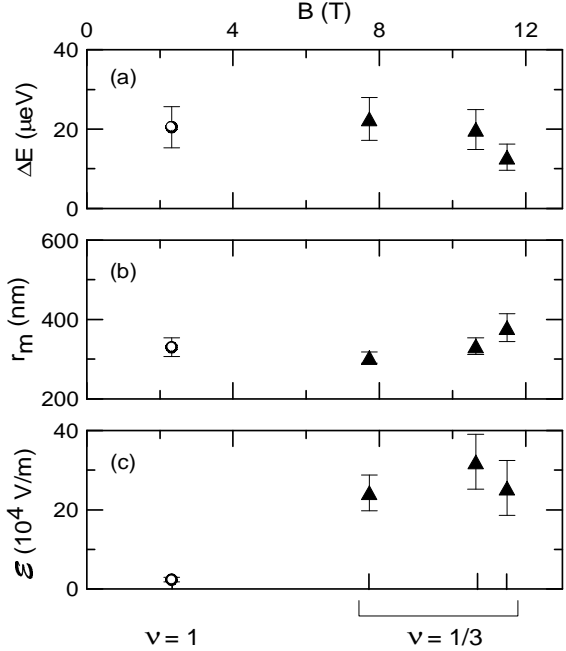,width=\linewidth,angle=0}}
\vspace{1in}
\caption{(a) The energy spacing $\Delta E = E_{m-1} - E_m$, (b) radius of the antidot $r_m$, and (c)  radial electric field ${\cal E}$, obtained as described in the text, as a function of magnetic field.}
\label{dee}
\end{figure}

To calculate ${\cal E}$ from $\alpha$ and $\beta$, from the Aharonov-Bohm quantization condition for a circular anidot-bound state with radius $r_m$, $\pi r_m^2B = m \phi_0$, we derive:

\begin{eqnarray}
\alpha &=& \frac{ {\cal E}\phi_0 r_m C_{ins}}{2B} , \\
\beta &=& \frac{q {\cal E} r_m}{2B}.
\label{micro}
\end{eqnarray} 
Since the values of $r_m$, Fig. 5 (b), can be determined independently from the period of RT conductance peaks,\cite{gold,peri} both Eqs. (5) and (6) can be used to calculate ${\cal E}$, which is shown in Fig. 5(c).  The value of ${\cal E} \sim 2 \times 10^4$ V/m for $\nu=1$ is consistent with estimates for quantum dots in integer QH-regime,\cite{mceu} although it is an order of magnitude higher for $\nu=1/3$. Apparently, one factor of 3 is needed to account for charge $e/3$ particles, and another factor of $\approx$ 3 comes from the ratio of magnetic fields. Finally, the drift velocity of edge excitations, $v_e = {\cal E}/B$ in our sample can be determined; we obtain $v_e = 1 \times 10^4$ m/s for $\nu=1$ and $v_e = 3 \times 10^4$ m/s for $\nu=1/3$. These values are somewhat lower than $1 \times 10^5$ m/s often used in theoretical estimates.\cite{chamon}

We would like to thank D. V. Averin, J. K. Jain, B. I. Halperin,  and S. A. Kivelson for interesting discussions and B. Su for help in sample fabrication. I. J. M. thanks Finnish Cultural Foundation for partial financial support. This work was supported in part by NSF under grant No. DMR-9629851.
%
%
%
%

\begin{references}
\bibitem{nuB} In the experiment $\nu$ in the constriction is smaller than $\nu_B$, the filling factor in the bulk, as discussed in Ref. 1.
\bibitem{Lutt} As has been shown in Fig. 3 of Ref. 4, the difference between Fermi-Dirac and Luttinger line shapes is less than 1 percent, and is not observable under our experimental conditions; thus, use of a specific line shape for data analysis does not affect results of this work.
\bibitem{sample} The sample was optimized for the fractional, rather than the integer QH regime.
\bibitem{Ref 10} $q =\phi_0 C_{ins} (\Delta V_{BG}/ \Delta  B)$ was first used in Ref. 1 to measure charge of the tunneling particles $q$.
\end{references}
\bibliographystyle{prsty}


\end{document}